\documentclass[letterpaper,twocolumn,10pt]{article}
\usepackage{usenix}

\usepackage{cite}
\usepackage{amsmath,amssymb,amsfonts}
\usepackage{textcomp}
\usepackage{xcolor}
\usepackage{float}
\usepackage{enumitem}
\usepackage[linesnumbered,ruled,vlined]{algorithm2e}
\usepackage{subcaption}
\usepackage{tabularx}
\usepackage{hyperref} 
\usepackage{array}
\usepackage{booktabs}
\usepackage{comment}
\usepackage{multirow}

\usepackage{listings}
\usepackage{xcolor}

\lstdefinelanguage{yaml}{
  keywords={true,false,null,y,n},
  keywordstyle=\color{blue}\bfseries,
  basicstyle=\ttfamily\small,
  sensitive=false,
  comment=[l]{\#},
  commentstyle=\color{gray}\ttfamily,
  stringstyle=\color{blue}\ttfamily,
  morestring=[b]",
  morestring=[b]'
}

\usepackage[pdftex]{graphicx}
\DeclareGraphicsExtensions{.pdf,.jpeg,.png}

\usepackage{tikz}
\usepackage{amsmath}

\usepackage{xurl}

\newcommand{\pName}{PrivLess\xspace}

\newcommand{\pTitle}{Permissions on the Loose: 
Measuring Overprivilege \\ 
in Real-World Serverless Applications}

\newcommand{\permInferenceSec}{Permission Inference\xspace}
\newcommand{\permEngine}{Permission Inference Engine\xspace}

\newcommand{\resGenSec}{Resource Generalization\xspace}
\newcommand{\resGeneralizer}{Resource Generalizer\xspace}

\newcommand{\polCompSec}{Policy Compilation\xspace}
\newcommand{\polCompiler}{Policy Compiler\xspace}

\newcommand{\shortsectionBf}[1]{\addvspace{3pt}\noindent {\bf #1}}
\usepackage{tablefootnote}
\usepackage{fancyvrb}

\providecommand{\Pass}{\ensuremath{\checkmark}}
\providecommand{\Fail}{\ensuremath{\times}}
\providecommand{\Blocked}{\ensuremath{\triangle}}
\providecommand{\NA}{\textemdash}

\def\BibTeX{{\rm B\kern-.05em{\sc i\kern-.025em b}\kern-.08em
    T\kern-.1667em\lower.7ex\hbox{E}\kern-.125emX}}

\begin{document}

\date{}

\title{\Large \bf \pTitle}

\author{
{\rm Elvis Yeboah-Duako}\\
University of Massachusetts, Amherst \\
eyeboahduako@umass.edu
\and
{\rm Pubali Datta}\\
University of Massachusetts, Amherst \\
pdatta@umass.edu
} 

\maketitle

\begin{abstract}
Serverless computing has seen rapid adoption in cloud deployments, yet the security implications of its service-oriented programming model remain poorly understood. Distributed, modular, and heterogeneous applications complicate the specification of precise security policies. Role-based access control solutions such as Identity and Access Management (IAM) already exhibit pervasive misconfiguration problems, and the multiplicity of functions, services, and resources in serverless applications, together with frequent permission model changes by cloud providers, greatly increases the likelihood of policy misconfigurations. Consequently, policies are often overprivileged, thereby enlarging the attack surface and exposing sensitive cloud resources to compromise.

We present a large-scale measurement study of overprivilege in real-world serverless applications, analyzing a curated dataset of 789 AWS Lambda applications comprised of 1,293 functions.
To enable this study, we develop PrivLess, a static policy analysis framework that extracts function-to-resource interactions from application source code, derives an interaction-permission mapping, and reconciles inferred interactions with declared policies to quantify overprivilege. Our measurement reveals that overprivilege is systemic and severe across the serverless ecosystem: 47.7\% of applications carry excess permissions with a significant privilege reduction potential of 99.65\%. Applications with wildcard-defined permissions exhibited an average overprivilege ratio 274$\times$ higher than those without. More critically, the excess permissions enable concrete attack vectors: 18.8\% of applications hold unnecessary Privilege Escalation capabilities, and 12 applications had Defense Evasion permissions they did not need.


\end{abstract}


\section{Introduction}
Identity and Access Management (IAM) frameworks (e.g., AWS IAM~\cite{amazonIdentityAccess}, Google IAM~\cite{googleIAM}, Microsoft IAM~\cite{microsoftrbac}) enable developers to specify security policies for cloud applications. However, writing correct policies remains difficult, and developers frequently introduce misconfigurations and excessive privileges. This problem is exacerbated in serverless environments, where modular, event-driven, stateless functions~\cite{PlanetOfServerless} must access many auxiliary services (e.g., data stores, messaging queues, logging), that makes least-privilege specification harder and expands the attack surface~\cite{Marin2022}. Over-privileged policies have contributed to high-impact breaches (e.g., SolarWinds~\cite{securityboulevardCloudInfrastructure}) and widespread incidents, with roughly 82\% of enterprises reporting security problems from misconfigured policies~\cite{cloud-report, MisconfigurationsDrive} and aggregate financial losses exceeding USD 5 trillion~\cite{fivemilliondollarloss}.

 Cloud platforms offer hundreds of services, each with numerous actions and resources requiring precise permission specifications. The ever-evolving nature of IAM permissions framework~\cite{amazonStrategiesAchieving,parisel2025quantifyingazurerbacwildcard}, opaque service dependencies, and inadequate documentation for correlating policy components~\cite{least-privilege-calls} complicate correct policy definition. Moreover, in serverless applications, each function acts as a distinct security principal requiring specific authorizations. This function-level granularity compounds policy complexity even further. Consequently, developers resort to coarse-grained policies with wildcards to expedite deployment. Moreover, even initially secure policies can become overprivileged as cloud providers introduce finer-grained permissions for their services over time. Updating application policies to adopt these refined permissions is labor-intensive and often neglected, leaving existing policies with excessive privileges ~\cite{forbesCloudMisconfigurations, hbrDataBreaches}.

 Given the severity of this problem, we set out, in this work, to measure how overprivileged serverless cloud applications are in the wild. To this end, we sampled and analyzed 789 real-world applications with a total of 1,293 serverless functions, from our curated dataset of 3,363 serverless applications sourced from open-source repositories. The sample size comprised 622 JavaScript/TypeScript, 151 Python, and 16 Go applications. We build \pName, a policy analysis framework to quantify overprivilege in serverless application policies before deployment. \pName employs static analysis to extract all service and resource interactions from function source code, and automatically generates a comprehensive resource-action-permission mapping from current cloud provider documentation and resources. \pName further reconciles discovered interactions with required permissions to precisely characterize overprivilege in developer-defined policies. We implement \pName for the AWS Lambda ecosystem and evaluate it on our curated dataset.  

To the best of our knowledge, this is the first study to systematically measure overprivilege in real-world serverless cloud applications through systematic code-to-policy analysis. Our analysis reveals pervasive and systemic overprivilege, with applications granted an average of 418.75 permissions while requiring only 1.55.
Applications with wildcards show a 99.02\% privilege reduction potential, with 69.2\% of these excess permissions granted on critical data services (e.g., S3, DynamoDB). Policy boundaries further shape the risks as function-specific policies yield a 3.60$\times$ overprivilege ratio versus 95.94$\times$ for global policies (one policy for all functions)
More critically, the excess permissions enable concrete attack vectors: 18.8\% of applications hold unnecessary Privilege Escalation capabilities, and 12 applications had Defense Evasion permissions they did not need.

These findings complement and inform existing work on policy analysis. The industry has introduced tools such as AWS IAM Access Analyzer~\cite{iam-access-analyzer} and Google Security Command Center~\cite{googleSecurityCommand} to help minimize policy misconfigurations. However, these tools operate on deployed configurations, identifying potential issues only after excessive privileges have been granted to cloud resources. In the research community, Valve~\cite{Valve} and Trapeze~\cite{Trapeze} track sensitive data flows but do not directly enforce permission boundaries. Policy generation tools \cite{GRASP, autoarmor, willIAm, DAntoni2024, gupta2024growlithe} infer access policies from execution logs, runtime behavior, or developer annotations, requiring either significant manual effort or live deployment prior to generating recommendations. By providing the first  characterization of overprivilege at scale, our measurement study establishes a concrete empirical foundation to guide the development of more effective, development-time least-privilege enforcement tools.
 
In summary, our contributions are as follows:
\begin{itemize}
     \item We conduct a large-scale empirical analysis of serverless applications and provide new insights into common IAM policy misconfigurations in serverless cloud applications. We quantify the extent of over-privilege and highlight opportunities to reduce it. 
     \item We create a dataset of 3,363 real-world serverless applications (1908 JavaScript/TypeScript, 400 Python, 64 Go, and others) to support reproducible research in serverless cloud security.
     \item We design a policy analysis framework for automatically reasoning about an optimal permission set needed to balance security and code functionality.

\end{itemize}
\section{Background and Motivation}
\label{sec:background}

IAM policies govern access to resources by specifying allow or deny rules for specific actions and principals. An IAM policy statement comprises four main components (see Listing~\ref{lst:iam-example}): (1) a security principal -- the entity being authorized; (2) a resource -- the system, service, or data to be accessed; (3) actions -- the operations the principal can perform on the resource; and (4) an effect -- either allow or deny. In serverless computing, each Lambda function serves as a security principal and must be granted the necessary permissions to access protected resources.
\begin{lstlisting}[language=yaml, caption={Example IAM Policy statement}, label={lst:iam-example}]
funcA:
   iamRoleStatements:
      - Effect: Allow
        Action:
           - "dynamodb:Get*"
        Resource: "accounts-table"
\end{lstlisting}

\subsubsection{IAM Policy Definition Types} 
In serverless environments, IAM policies can be defined globally so that all functions in an application inherit the same level of authorization (\emph{Global policies}). Such policies can be inherently overly permissive, as not all functions require the same privileges. On the other hand, IAM policies can be defined with custom authorizations for each function (\emph{Function-based policies}). While this is the recommended way to define policies, they can still be overly permissive when a function's permission set exceeds the required level. Moreover, when many individual functions are considered, defining policies for them becomes harder to reason about across varying levels of function complexity. As a result, developers sometimes favor global security policies, accepting the associated security vulnerabilities and risks \cite{globalpolicyinsecure}.

\subsubsection{Wildcard usage in IAM policies} 
Wildcards match multiple actions or resources by using pattern matching. In Listing~\ref{lst:iam-example}, \texttt{dynamodb:Get*} is a \emph{prefix} wildcard that allows a function to perform all DynamoDB actions beginning with \texttt{Get}. \emph{Service-level} wildcards are more permissive: \texttt{s3:*} grants all 106 AWS S3 actions to a function, yet most applications require only a small subset. The most permissive wildcard is the \emph{full} wildcard (\texttt{*}), which grants all possible actions on all AWS services—a common misconfiguration in practice. Similar wildcard patterns apply to resources where resource identifiers can be specified as full wildcards (all resources in the tenant), service wildcards (all resources of a given service), or prefix wildcards (resources matching a specific path). These resource wildcards are equally problematic, as they grant access to resources most functions do not need access to.

\subsubsection{Managed Policies}
Managed policies are predefined security policies designed for common use cases to simplify permission management. They assist developers with limited expertise in policy definition by bundling necessary permissions for routine tasks. While managed policies represent an improvement over wildcards, they frequently grant permissions broader than an application requires, potentially introducing security risks. Moreover, because developers do not control managed policies, updates by cloud providers or administrators can have unintended consequences. A notable example occurred in December 2021, when AWS inadvertently added the \texttt{s3:GetObject} permission to the \texttt{AWSSupportServiceRolePolicy} managed policy, despite its intention to provide only metadata visibility~\cite{aws-mp}. This incident illustrates how changes to managed policies can be difficult to track and difficult to reason about, emphasizing the need for tools that continuously audit actual permission requirements rather than relying on static, cloud-provider-defined policies.

\subsubsection{Motivating Example}
\begin{figure}
    \centering
    \includegraphics[width=\linewidth]{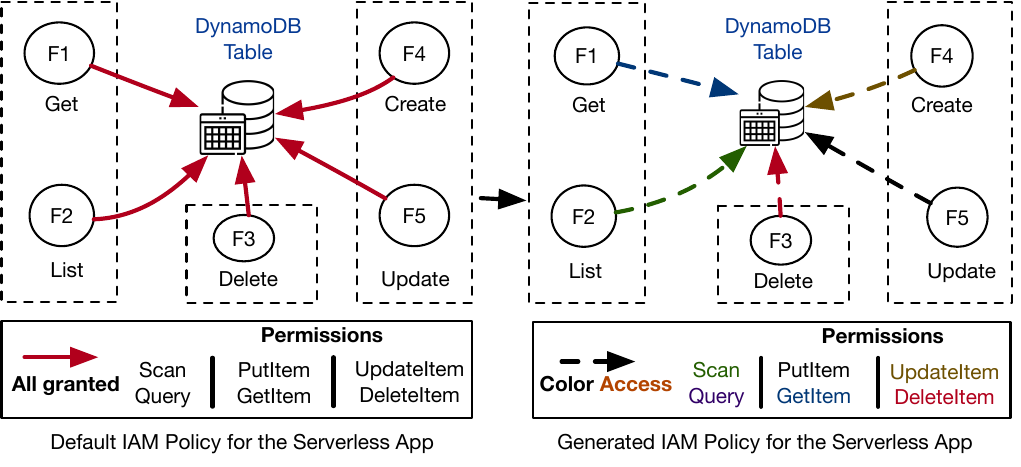}
    \caption{Overprivilege widens the blast radius by granted functions permissions they do not require.}
    \label{fig:motivation}
\end{figure}

To illustrate the overprivilege problem concretely, consider a representative serverless application from our dataset that implements a task-management API.\footnote{App anonymized for responsible disclosure} The application comprises five serverless functions that interact with a single DynamoDB table. It has a global policy with 6 permissions: \texttt{dynamodb:Query}, \texttt{dynamodb:Scan}, \texttt{dynamodb:GetItem}, \texttt{dynamodb:PutItem}, \texttt{dynamodb:UpdateItem}, and \texttt{dynamodb:DeleteItem}, as shown in Figure\ref{fig:motivation}. However, analyzing the application code reveals that each function required only a single permission. For example, the \texttt{List} function required only \texttt{dynamodb:Scan} to retrieve tasks, yet it was granted \texttt{dynamodb:DeleteItem} and \texttt{dynamodb:UpdateItem}. If this function is compromised through a code injection vulnerability or malicious dependency, an attacker could exploit these excess permissions to delete entire tables or corrupt all task records, causing immediate data loss and denial of service. Our analysis, shown in Figure~\ref{fig:motivation}, identifies this over-privilege and recommends a reduced permission set needed by each function. Applying these refined permissions achieves an 83.3\% reduction in privileges, substantially limiting an attacker's ability to cause damage while preserving all legitimate functionality.



\section{Threat Model and Assumptions}
\label{sec:threats}


We consider an adversary whose 
objective is to exploit overprivileged serverless function's IAM execution role to access sensitive cloud resources reachable from the function, and extend their reach beyond the compromised function. With excess permissions, an attacker can perform many unauthorized actions including the following:
(i)\emph{exfiltrate data} from storage resources that the function was never intended to access;
(ii)\emph{move laterally} to services unrelated to the function's purpose;
(iii)\emph{escalate privileges} (e.g., in AWS, via actions such as \texttt{iam:PassRole} or \texttt{sts:AssumeRole}) to acquire broader permissions; and
(iv)\emph{establish persistence} by creating new resources or modifying existing policies.

We assume that developer-defined policies are functionally correct and authorize the function to perform its intended operations. However, we make no assumption about the minimality of these policies; they may grant significantly more permissions than necessary for operational purposes. This allows \pName to use developer policies as an authoritative reference for legitimate function behavior while still identifying and quantifying over-privilege.

We do not consider malicious developers (insider threat), compromise of cloud infrastructure, or resource-based policies (e.g., Lambda resource policies) that grant access independently of the execution role.

\section{Serverless Applications Dataset}
\label{sec:datasets}
To the best of our knowledge, there is only one publicly available dataset
of serverless applications \cite{GRASP}. However, that dataset was curated in
2021, and the serverless ecosystem on GitHub has grown substantially since then. To address this gap, we curated a new 
dataset as described below. We focus on the \textit{Serverless Framework}~\cite{ServerlessFramework} because it is a widely adopted serverless deployment tool, and has standardized configuration that enables consistent large-scale static analysis without external policy lookups.

\shortsectionBf{Data Collection.}
We curated our dataset using a multi-stage filtering of GitHub repositories. First, we queried GitHub's Search API for repositories explicitly tagged with "serverless" in their metadata, which returned 167,290 repositories. To filter out experimental or incomplete projects, we applied a minimum threshold of 10 GitHub stars and excluded repositories created before 2014, when serverless computing emerged. This initial filtering reduced the dataset to 5,227 repositories, of which 69\% had been updated within the past year, indicating active maintenance. The collection phase was conducted over three days (October 17–19, 2025).

Next, we examined the cloned repositories to identify those built with the Serverless Framework~\cite{ServerlessFramework} by checking for the presence of a \texttt{serverless.y\{a\}ml} configuration file, which is required to use the framework. This filtering yielded 1,273 repositories.

Finally, we accounted for monorepository practices (large repositories containing multiple independent applications) by extracting each serverless application as a separate entity and excluding non-serverless applications. This extraction process identified 3,714 applications from the 1,273 repositories. Of these, 351 applications (9.45\%) were excluded due to parsing failures: invalid configuration syntax (82), unimplemented intrinsic function calls (17) such as \texttt{Fn::FindInMap}, \texttt{Fn::Sub}, and \texttt{Fn::ImportValue} that required live API access; conversion errors (89) from unhandled event types and type mismatches; and missing provider or runtime sections (163) required for analysis. Our final dataset contained 3,363 serverless applications.

\begin{table}[H]
\centering
\caption{Serverless Projects by Cloud Provider}
\label{tab:provider-distribution}
\setlength{\tabcolsep}{6pt}
\renewcommand{\arraystretch}{1.2}

\begin{tabular}{lr}
\hline
\textbf{Cloud Provider} & \textbf{App Count (\%)} \\
\hline
AWS       & 2553 (75.91\%) \\
Scaleway  & 41 (1.22\%)    \\
Azure     & 39 (1.16\%)    \\
Google    & 29 (0.86\%)    \\
Others (OpenWhisk, Cloudflare, etc)     & 701 (20.85\%)  \\
\hline
\textbf{Total} & \textbf{3363 (100\%)} \\
\hline
\end{tabular}

\end{table}

\shortsectionBf{Cloud Provider Distribution.}
\label{sec:provider-dist}
As shown in Table~\ref{tab:provider-distribution}, the dataset is heavily skewed toward AWS. Amazon Web Services (AWS) dominates with 2,553 applications (75.91\%), while Microsoft Azure Functions and Google Cloud Functions account for 39 (1.16\%) and 29 (0.86\%), respectively. The remaining 20.85\% of applications use alternative providers, including OpenWhisk, Tencent, Cloudflare, Kubeless, and OpenFaaS.

\section{Methodology}

\begin{figure}[t]
    \centering
    \includegraphics[width=\linewidth]{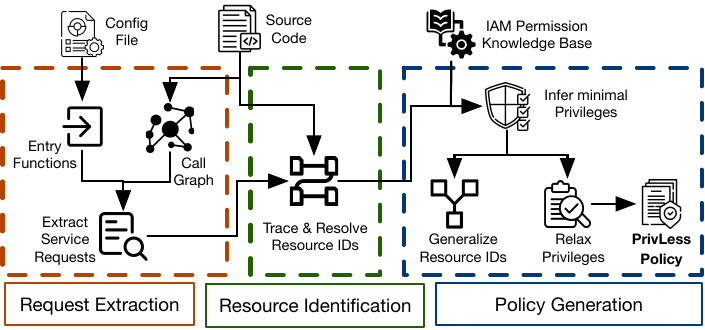}
    \caption{Overview of \pName's source code analysis for minimal policy generation}
    \label{fig:privLess-sys}
\end{figure}

We designed \pName, a policy overprivilege analysis pipeline that quantifies and reduces overprivilege in serverless applications. At its core, \pName analyzes application source code to extract all function-to-cloud-service interactions, generates minimal permission sets based on the observed interactions, and reconciles them with developer-defined IAM policies to identify excess permissions. It then synthesizes refined policies containing only necessary privileges (Figure~\ref{fig:privLess-sys}). To enable this analysis, we first construct an offline curated IAM Knowledge base that maps each API endpoint (SDK call) to its minimum required permissions.

\subsection{IAM Permission Knowledge Base}
\label{subsec:kb}
Determining whether a serverless function holds excess permissions requires knowing the minimum permission(s) each service request actually needs. While cloud providers document which permissions are required for specific actions, this information is fragmented across multiple sources(e.g., SDK documentation, API references, managed policies) and is often incomplete or difficult to reconcile. To address this, we construct a curated IAM permission Knowledge Base by running six complementary extraction strategies over the provider's SDK artifacts and public documentation, and then converging their outputs into an action-to-permission mapping.

\begin{table*}[t]
\centering
\caption{IAM permission extraction strategies (cloud-agnostic).}
\label{tab:strategies}
\footnotesize
\setlength{\tabcolsep}{5pt}

\begin{tabular}{c l p{7cm} p{5.5cm}}
\toprule
\textbf{ID} & \textbf{Strategy} & \textbf{Signal (Extraction method)} & \textbf{Source (Sample per provider)} \\
\midrule

\#S1 & Direct mapping &
Near-universal naming convention for mapping of cloud SDK methods to their corresponding permissions or actions (e.g, \texttt{s3.getObject() -> s3:GetObject)} &
AWS: AWS SDK (e.g., botocore service-2.json)~\cite{aws:botocore}; Google Cloud: API client libraries~\cite{gcp:client-lib}; Azure: Azure SDK~\cite{azure:sdk} \\

\#S2 & Regex mining &
Conditional permissions embedded in prose documentation in code, keyed to the parameter that triggers them &
AWS: botocore doc strings~\cite{aws:botocore}; Google Cloud: Cloud API documentation~\cite{gcp:docs}; Azure: API reference~\cite{azure:docs} \\

\#S3 & Semantic annotations &
Machine-readable permission annotations authored by the cloud provider's service team, providing cross-service dependency signals &
AWS: AWS SDK models (Smithy)~\cite{aws:smithy}; Google Cloud: API definitions~\cite{gcp:api-def}; Azure: OpenAPI definitions~\cite{azure:openapi} \\

\#S4 & Heuristic rules &
Structured rule table encoding cross-service patterns (encryption, identity delegation, networking, container registries) that fire on matching parameter names &
Parameter-shape analysis from service SDK specifications (provider-agnostic) \\

\#S5 & Authorization reference &
Available cloud documentation. Ground-truth enumeration of all valid actions/permissions mapping; serves as both data source and validation corpus &
AWS: IAM Authorization Reference~\cite{aws:sar}; Google Cloud: IAM Roles Reference~\cite{gcp:iam-roles}; Azure: RBAC Built-in Roles~\cite{azure:rbac-roles} \\

\#S6 & Provider managed policies &
Service-to-service dependency index derived from cloud-provider-managed and service-linked policies; flags implausible cross-service entries &
AWS: Managed policies corpus~\cite{mamip}; Google Cloud: Predefined roles~\cite{gcp:predefined-roles}; Azure: Built-in roles~\cite{azure:builtin-roles} \\

\bottomrule
\end{tabular}
\end{table*}

\paragraph{Extraction strategies.}
We employ six complementary strategies (\#S1–Direct mapping, \#S2–Regex mining, \#S3–Semantic annotations, \#S4–Heuristic rules, \#S5–Authorization reference, \#S6–Managed policies) to construct the curated Knowledge Base. Table~\ref{tab:strategies} summarizes each strategy's signal, extraction method, and data sources. \#S1 and \#S5 form the backbone: \#S1 derives a primary permission for every operation using the cloud provider's naming convention, while \#S5 provides the authoritative enumeration of valid IAM actions. Any permission \#S1 derives that is absent from \#S5 is flagged \texttt{not\_in\_docs}. \#S3 contributes the most reliable cross-service signals, as its annotations are embedded in the SDK's formal service model. \#S2 and \#S4 surface conditional and implicit permissions not found in formal documentation, prioritizing recall over precision. Finally, \#S6 acts as a statistical validator: cross-service permissions whose caller-to-callee service pair never appears in the provider's managed or predefined policies are down-weighted as likely false positives.

\vspace{0.5em}
\begin{algorithm}[htbp]
\caption{IAM Knowledge Base Convergence}
\label{alg:convergence}

\DontPrintSemicolon
\SetAlgoLined

\KwIn{
    $\mathit{hits}$: strategy $\rightarrow$ operation data \\
    $\mathit{valid}$: authoritative reference action set
}

\KwOut{
    $\mathit{primary}$: list 
    $\langle \mathit{permission},\, \mathit{conf} \rangle$ pairs
}


$\sigma \gets \{\}$\;

\ForEach{strategy $s$ with primary permission $p$ in $\mathit{hits}[s]$}{
    $\sigma[p] \mathrel{+}= w_s$\;
}

$p^* \gets \arg\max_p \sigma[p]$\;

\lIf{$\sigma[p^*] \geq 4$}{
    $\mathit{conf} \gets \textsc{high}$
}
\lElseIf{$\sigma[p^*] \geq 2$}{
    $\mathit{conf} \gets \textsc{medium}$
}
\lElse{
    $\mathit{conf} \gets \textsc{low}$
}

\Return{$[\langle p^*,\, \mathit{conf} \rangle]$}\;

\end{algorithm}
\vspace{0.5em}

\paragraph{Convergence and confidence.}
The outputs of the six strategies are merged per $\langle\mathit{service}, \mathit{operation}\rangle$ pair via weighted voting with weights $w_{\#S5}\!=\!w_{\#S3}\!=\!3$, $w_{\#S1}\!=\!2$, $w_{\#S2}\!=\!w_{\#S4}\!=\!1$. Algorithm~\ref{alg:convergence} details the core convergence process. For each permission candidate $p$, we accumulate the weights of all strategies that identified it, selecting the highest-scored candidate $p^*$ and assigning confidence based on the cumulative score: $\sigma \geq 4 \Rightarrow \textsc{high}$; $\sigma \geq 2 \Rightarrow \textsc{medium}$; otherwise $\textsc{low}$. Post-convergence, quality flags may be applied to annotate entries with known limitations: for example, services whose IAM naming is incompatible with the 1:1 permission-to-operation mapping, or permissions derived by strategies but absent from the authoritative reference source. These flags inform downstream users of the knowledge base about entry confidence and applicability. Conditional and cross-service permissions are converged using the same weighted scoring and confidence assignment process.

\paragraph{Runtime resolution.}
At analysis time the registry is queried with a $\langle\mathit{service},\mathit{action},\mathit{params}\rangle$ triple. Conditional permissions are included only when their trigger parameter is present in the call, narrowing the permission set to what the specific
invocation actually requires. 
This results in a per-function minimal permission set $\mathcal{R}_f$ (defined formally in Section~\ref{sec:perm-inference}) used to compute over-privilege as $\mathcal{G}_f \setminus \mathcal{R}_f$, where $\mathcal{G}_f$ is the baseline permission set granted to function $f$'s execution role
by the developer.

\subsection{Source-Code Analysis}

\subsubsection{Request extraction}
With the IAM knowledge base setup, we analyze application source code statically to identify all interactions between defined serverless functions and cloud services. A function $f_i$ may invoke one or more service API endpoints (wrapped by the cloud provider's SDKs). We refer to the set of statically determined API calls as service requests, $R$. Each request $r_i = (f_i, a_i, q_i) \in R$ consists of a service action $a_i$ (e.g., \texttt{s3.getObject}) and a set of parameters $q_i$ (e.g., \texttt{(Bucket=<bucket>, Key=<key>)}).

First, given a serverless app, we parse the configuration file to identify the entry functions. 
Then, for each function, we construct a call graph to identify dependent functions and modules, ensuring that privileges required by such functions or inter-procedural calls are properly attributed to the entry function. We then trace the instantiations and use of a cloud provider's imported SDK in the code to extract all the service requests, $R$ invoked by the functions. 
$R$ is passed to the next phase, which infers the resource objects targeted by these requests.

\subsubsection{Resource Identification}
The parameters, $q_i$, in a given service request, $r_i$, may contain resource identifiers that the request targets. Such resource IDs (e.g., S3 bucket names or DynamoDB table names) are essential for ensuring that access controls are properly scoped to the intended resources. While many resource values are dynamically provisioned, we use data flow analysis~\cite{kennedy1979survey} to trace parameter variables to their concrete values where statically determinable—from literal assignments, environment variables, or the serverless configuration file. Not all parameters specify a resource, so we curate a resource key mapping for all applicable services based on provider's documentation and attempt resource ID resolution only for parameters present in this mapping (e.g., \texttt{Bucket} for s3). Once a resource ID is successfully resolved, we replace the variable name in the initial request with the resolved ID. Otherwise, we retain the first declared or instantiated value which is typically an environment variable or declared variable that flows into the parameter. We argue that, although static analysis cannot capture all resource IDs due to their dynamic nature, by using the environment variable or declared variable, developers can tell which resources they intend to secure, making explicit specification straightforward. Moreover, by preserving unresolved environment variables in the policy, they resolve correctly to the intended resources at runtime.

\subsubsection{Policy Generation}
Given the updated requests, $R$, \pName generates the final IAM policy through four sequential steps: (1)~A \permEngine queries the knowledge base for each request to produce the per-function inferred permission set $\mathcal{R}_f$; (2)~the \resGeneralizer consolidates resource identifiers into prefix wildcards, producing generalized requests $\mathit{GR}$; (3)~a relaxation step refines $\mathcal{R}_f$ against the developer's declared policy, yielding the relaxed set $\mathcal{R}^{*}_f$; and (4)~the \polCompiler assembles $\mathit{GR}$ and $\mathcal{R}^{*}_f$ into the final policy $P_N$.

\shortsectionBf{(1) \permInferenceSec.}
\label{sec:perm-inference}
Given the set of requests, $R$, the \permEngine resolves the minimum IAM permissions required for each request by querying the knowledge base (Section~\ref{subsec:kb}). For each request $r_i = (f_i, a_i, q_i)$, the knowledge base is queried with the triple $\langle \mathit{service}, a_i, q_i \rangle$: it returns the primary IAM action and any parameter-triggered conditional permissions. For example, in AWS, \texttt{s3.getObject(..., VersionId)} resolves to \texttt{s3:GetObjectVersion} permission, and to \texttt{s3:GetObject} when the \texttt{VersionId} parameter is not present (i.e VersionId $\notin q_i$). Moreover, conditional cross-service permissions (e.g., \texttt{kms:GenerateDataKey} when an SSE-KMS key parameter is present) are included only when the triggering parameter is present in $q_i$. The union of permissions resolved across all $r_i \in R$ with $f_i = f$ constitutes the per-function inferred required set $\mathcal{R}_f$.

\shortsectionBf{(2) \resGenSec.}
Serverless applications often access multiple objects or resources from a shared location. For example, an application may request these files from an S3 bucket: \textit{bucket/images/image1.jpg}, \textit{bucket/images/image2.jpg}, \textit{bucket/images/image3.jpg}, and so on. Resource generalization summarizes these requests into a prefix wildcard that represents the shared directory: \textit{bucket/images/*}. This allows \pName to generate a single IAM policy statement that grants the applicable function the necessary permissions for all objects in the directory, rather than creating separate statements for each individual object—an approach that would result in a complex and unwieldy policy. Moreover, this ensures that the generated IAM policy can accommodate future requests to resources in the same path without modification.

The \resGeneralizer groups the requests, $R$, by function, $f_i$. We limit resource generalization to requests made by the same function to ensure that the generated IAM policy is specific to each function's operations and does not inadvertently grant other functions access to resources they do not require. For each group of requests, the \resGeneralizer examines the resource IDs in $q_i$ to identify and generate prefix wildcards for resources sharing a common path. The common prefix, $\mathit{cp}_i$, becomes the generalized resource identifier for all requests in that subset. Resources with unique paths are preserved without generalization. Finally, \pName outputs a set of generalized requests, $\mathit{GR}$.

\shortsectionBf{(3) Privilege Relaxation.}
\label{sec:relaxation}
Taking $\mathcal{R}_f$ (the per-function inferred required set from permission inference) and $\mathcal{G}_f$ (the developer-defined grant set for function $f$, parsed from the framework configuration) as inputs, this step addresses two limitations of purely static inference. First, $\mathcal{R}_f$ may contain permissions the developer never intended to grant. Second, implicit dependencies that are absent from explicit SDK calls (e.g., runtime-emitted log writes) may be missing from $\mathcal{R}_f$, causing legitimate permissions to be incorrectly flagged as over-privileged.\pName applies a two-pass relaxation step.

\textbf{Pass~1 (Bounding).} Let $\hat{\mathcal{G}}_f$ denote the wildcard-expanded developer grant set, where patterns such as \texttt{svc:*} or \texttt{svc:Put*} are treated as covering all matching actions. Any permission in $\mathcal{R}_f$ not covered by $\hat{\mathcal{G}}_f$ is evicted, yielding $\mathcal{R}_f \cap \hat{\mathcal{G}}_f$. This bounds the inferred set strictly by developer intent, given the assumption that developer-defined permissions are functional(Section \ref{sec:threats}). Any false-positive permissions generated by the knowledge base are thereby corrected.


\textbf{Pass~2 (Augmentation).} To recover implicit permissions that are opaque to static analysis and the IAM permission knowledge base, we treat each developer-defined permission $p \in \mathcal{G}_f \setminus \mathcal{R}_f$ as a candidate for inclusion in $\mathcal{R}_f$ if and only if all of the following conditions hold:

\begin{itemize}
\item \textit{The candidate's service is observed for the function.} Co-occurrence elsewhere in the dataset is insufficient, since jointly used services do not imply that all permissions of one are required whenever the other is invoked.

\item \textit{The candidate is not statically inferred for another function in the application.} If PrivLess detects the permission elsewhere in the same codebase, its absence here indicates that the function does not exercise, and not that the analysis missed it.

\item \textit{The candidate's access level is consistent with the function's observed access pattern.} A function exhibiting only read-level accesses cannot acquire a write-level permission through augmentation. If the candidate's access level is unknown, this condition is satisfied by default.
\end{itemize}

Admitted candidate permissions are combined with the permissions retained by Pass~1 to form the final relaxed set $\mathcal{R}^{*}_f$. Because all augmented permissions originate from the developer's grant, $\mathcal{R}^{*}_f \subseteq \hat{\mathcal{G}}_f$, preserving developer intent without broadening the original grant. 

\shortsectionBf{(4) \polCompSec.}
Finally, the \polCompiler takes $\mathit{GR}$ and the per-function relaxed permission sets $\mathcal{R}^{*}_f$ as inputs and assembles them into IAM policy statements $P_N$, grouping by function and then by resource. For each resource accessed by a function, a single policy statement is generated that assigns the corresponding permissions. For instance, In an AWS serverless application, if a function issues three requests $r_1, r_2, r_3$ where $r_1$ and $r_3$ are read and write requests to access an AWS S3 resource \textit{bucket/images/} and $r_2$ is a read request to resource \textit{bucket/videos/}, the \polCompiler generates:

\begin{Verbatim}[fontsize=\small]
iamRoleStatements:
    - Effect: "Allow"
      Action: ["s3:GetObject", "s3:PutObject"]
      Resource: "bucket/images/"
    - Effect: "Allow"
      Action: ["s3:GetObject"]
      Resource: "bucket/videos/"
\end{Verbatim}

The aggregated policy, $P_N$ can then be compared against the developer-defined baseline policy to quantify overprivilege.
\section{Implementation}
\label{sec:implementation}
We implemented a prototype of \pName in Python and integrated it with CodeQL~\cite{githubCodeQL} for static and data flow analysis of source code. CodeQL's semantic code analysis engine provides a rich query language that enables seamless retrieval of information from its relational database representation of application source code. 

\shortsectionBf{Measurement Scope.}
The dominance of AWS (75.91\% of applications) in our dataset (Section~\ref{sec:provider-dist})  aligns with its leading position in the serverless computing market~\cite{datadoghqStateServerless}. More importantly, a structural property of the Serverless Framework's AWS integration makes static over-privilege analysis tractable: developer-defined IAM permissions can be declared inline within the application configuration file (\texttt{serverless.yml}), making the developer-granted permission set $\mathcal{G}$ directly accessible to quantify overprivilege. In contrast, Google Cloud Functions and Azure Functions use fundamentally different permission models. Both providers bind permissions to service accounts or managed identities through out-of-band operations—\texttt{gcloud} CLI role bindings~\cite{gcp:functions-iam} or Azure RBAC assignments~\cite{azure:rbac}—that reside outside the application repository. Consequently, this measurement study focuses on AWS serverless applications, as developer-defined permissions are unavailable in source code for other providers, precluding over-privilege quantification.

\shortsectionBf{Supported Programming Languages.}
\label{sec:languages}
We implemented our analysis pipeline for the three most prevalent languages among AWS applications in our dataset (Table~\ref{tab:language-distribution}): JavaScript/TypeScript, Python, and Go, which collectively account for 86.2\% (2,244) of applications. This focus aligns with industry trends identifying these languages as the most popular for serverless workloads~\cite{datadoghqStateServerless, stackoverflowTechnology2024}. Our implementation uses language-specific CodeQL queries for semantic code analysis, while the core Python analysis pipeline remains language-agnostic. Support for additional languages (Java, C\#, Ruby, etc.) can be readily extended by writing analogous CodeQL queries for those languages.

\begin{table}[h]
\centering
\caption{Programming Languages in AWS Serverless Applications}
\label{tab:language-distribution}
\setlength{\tabcolsep}{6pt}
\renewcommand{\arraystretch}{1.2}
\begin{tabular}{lr}
\hline
\textbf{Language} & \textbf{App Count (\%)} \\
\hline
JavaScript/TypeScript & 1829 (71.6\%) \\
Python                & 363 (14.2\%) \\
Go                    & 52 (2.0\%) \\
Java                  & 40 (1.6\%) \\
Other                 & 269 (10.5\%) \\
\hline
\textbf{Total}        & \textbf{2553 (100\%)} \\
\hline
\end{tabular}
\end{table}

\shortsectionBf{Analyzed Applications and Functions.}
\label{sec:total_apps}
Of the 2,244 AWS serverless applications written in the evaluated programming languages, only 886 (39.48\%) contained at least one serverless function with an IAM policy defined, making them suitable for over-privilege analysis. However, 97 of these 886 applications (10.95\%) failed to complete the analysis pipeline due to CodeQL database creation failures, query execution timeouts, and incompatible Node/Python/Go versions. These applications were excluded from the study. The final analyzed dataset comprised 789 serverless applications with a total of 1,293 serverless functions. However, the full dataset of 3,363  multi-cloud-and-language serverless applications (as described in Section~\ref{sec:datasets}) would be made available to support varying research in serverless cloud security.
\section{Evaluation: Policy Analysis}
We run \pName against our dataset on a virtual machine with a 48-core AMD EPYC 7643 processor, 16 GB of memory, and 285 GB of disk space, running Ubuntu 22.04.5 LTS. Our evaluation has two main parts. First, we analyze developer-defined (baseline) policies to characterize permission-assignment patterns and quantify common IAM policy misconfigurations in the wild. Second, we use \pName to generate permissions inferred from observed code behavior and quantify overprivilege by comparing these reduced policies with the baseline.

\begin{figure*}
    \centering
    \includegraphics[width=\linewidth]{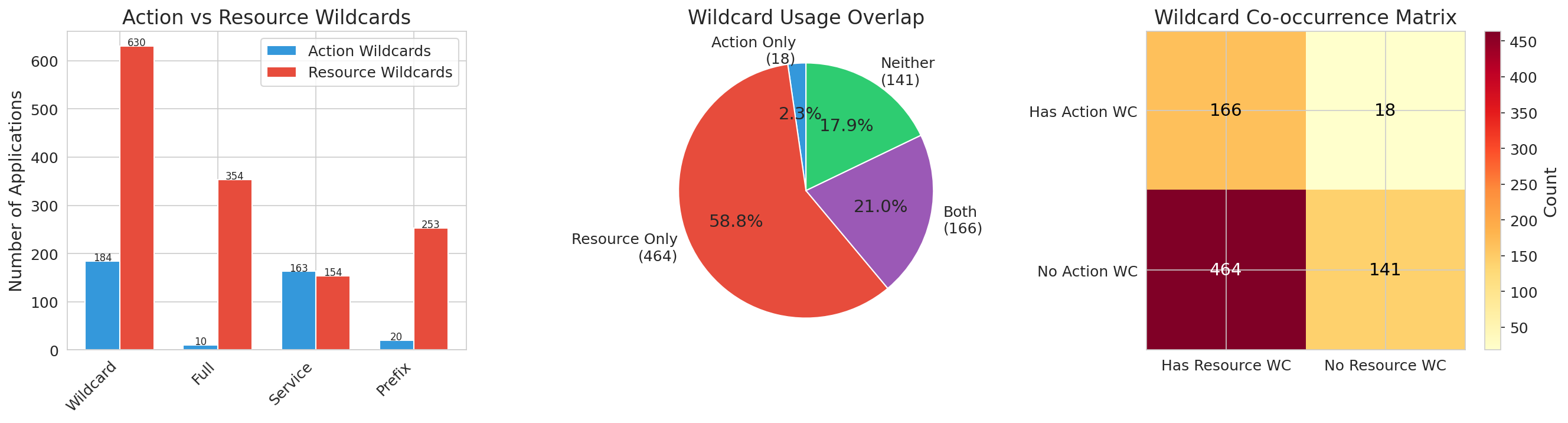}
    \caption{Wildcard distribution for permissions and resources across the dataset. 166 (21\%) applications had both permission and resource wildcards, indicating significant overprivilege.}
    \label{fig:wildcard_distribution}
\end{figure*}

\subsection{Baseline Policy Analysis}
\shortsectionBf{Policy Type Distribution.}
Across all 789 analyzed applications, global-only IAM policies dramatically dominate at 88.6\% (699 apps), revealing that developers overwhelmingly default to granting a single permission set across all functions, which inadvertently increases the blast radius of the application. This vulnerability was seen in breaches like the Capital One breach~\cite{capitalOne}, where a compromised function accessed S3 buckets it was not supposed to access. On the other hand, only 6.6\% (52 apps) followed the recommended approach for least-privilege enforcement by defining function-based IAM policies. We show in later analysis that even these scoped policies often are similarly overprivileged. Also, a small fraction (2.4\%, 19 apps) defined both global and function-based policies, which can be ideal for scenarios where all functions require similar access on some resources, in addition to other scoped permissions. Finally, there were some 19 apps (2.4\%) that relied on CloudFormation default roles outside the application's control, making them invisible to our static analysis pipeline.

\begin{table}[h]
\centering
\caption{Wildcard Usage in Permissions and Resources}
\label{tab:wildcard-distribution}
\setlength{\tabcolsep}{6pt}
\renewcommand{\arraystretch}{1.2}
\begin{tabular}{lcc}
\hline
\textbf{Wildcard Type} & \textbf{Permissions} & \textbf{Resources} \\
\hline
Full    & 10 (5.4\%)    & 354 (56.2\%) \\
Service   & 184 (88.6\%) & 154 (24.4\%) \\
Prefix    & 15 (10.9\%)  & 253 (40.2\%) \\
\hline
\textbf{Total} & \textbf{184} & \textbf{630} \\
\hline
\end{tabular}
\end{table}

\shortsectionBf{Action Wildcard Distribution.}
Wildcard usage in permissions represents one of the most serious forms of over-privilege in IAM policies. In our dataset, 23.3\% (184 apps) included at least one wildcard permission, indicating widespread misuse of this dangerous pattern. Table~\ref{tab:wildcard-distribution} breaks down the observed wildcard usage distribution.

The most critical findings are the 10 applications (5.4\% of 184) using full wildcards (\texttt{*}), which grant unrestricted access to all AWS services and resources. Compromise of any function in these applications amounts to near-complete account takeover: an attacker could create IAM users, exfiltrate data across all services, or establish persistent backdoors with no technical constraints. This represents a fundamental break of the security model.

More prevalent are the 163 applications (88.6\%) employing service-level wildcards (\texttt{<service>:*}), which allows all actions within specific services (e.g., S3, DynamoDB). While relatively limited in scope, these still violate least-privilege principles by granting unnecessary permissions within each service; a function that only reads S3 configuration should never have permissions to delete buckets, modify ACLs, or make data public. 

Finally, 20 applications (10.9\%) used prefix wildcards (\texttt{<service>:<prefix>*}), such as \texttt{dynamodb:Batch*}, granting groups of related actions when functions typically need only one or two specific operations. While the blast radius is smaller, this pattern still demonstrates inadequate permission governance.


\shortsectionBf{Resource Wildcard Distribution.} Resource wildcards compound the over-privilege problem. In our study, 79.8\% (630 apps) used wildcards in their resource definitions, demonstrating how action-level permissions are amplified by overly broad resource scopes (Table~\ref{tab:wildcard-distribution}). Most critically, 354 applications (56.2\% of 630) used full resource wildcards (\texttt{Resource: "*"}), allowing functions to access any resource in the account regardless of ownership or purpose. The consequences are severe. A policy granting \texttt{s3:GetObject} on \texttt{Resource: "*"} permits functions to read every S3 bucket in the account, including those containing credentials, application secrets, customer PII, and data from other applications. A single compromised function can thus trigger account-wide data exfiltration. This exact pattern enabled the Capital One breach~\cite{capitalOne}, where wildcard resources exposed more than 700 S3 buckets and sensitive customer data.

Service-scoped resource wildcards appear in 24.4\% (154 apps), such as \texttt{Resource: "arn:aws:dynamodb:region:account:table/*"}. While nominally limiting access to dynamoDB service resources, these still grant unrestricted access to all dynamoDB tables in a tenant's directory, including production databases and backups from other applications.

Prefix wildcards, used by 40.2\% (253 apps), represent a more disciplined approach. For example, \texttt{Resource: "arn:aws:dynamodb:region:account:table/app-*"} scopes access to tables matching a naming convention. However, these can still enable over-privilege if the prefix patterns are too broad or inadvertently span unrelated resources. Even with careful naming, prefix wildcards remain a compromise between convenience and principle.

\begin{figure}[t]
    \centering
    \includegraphics[width=0.8\linewidth]{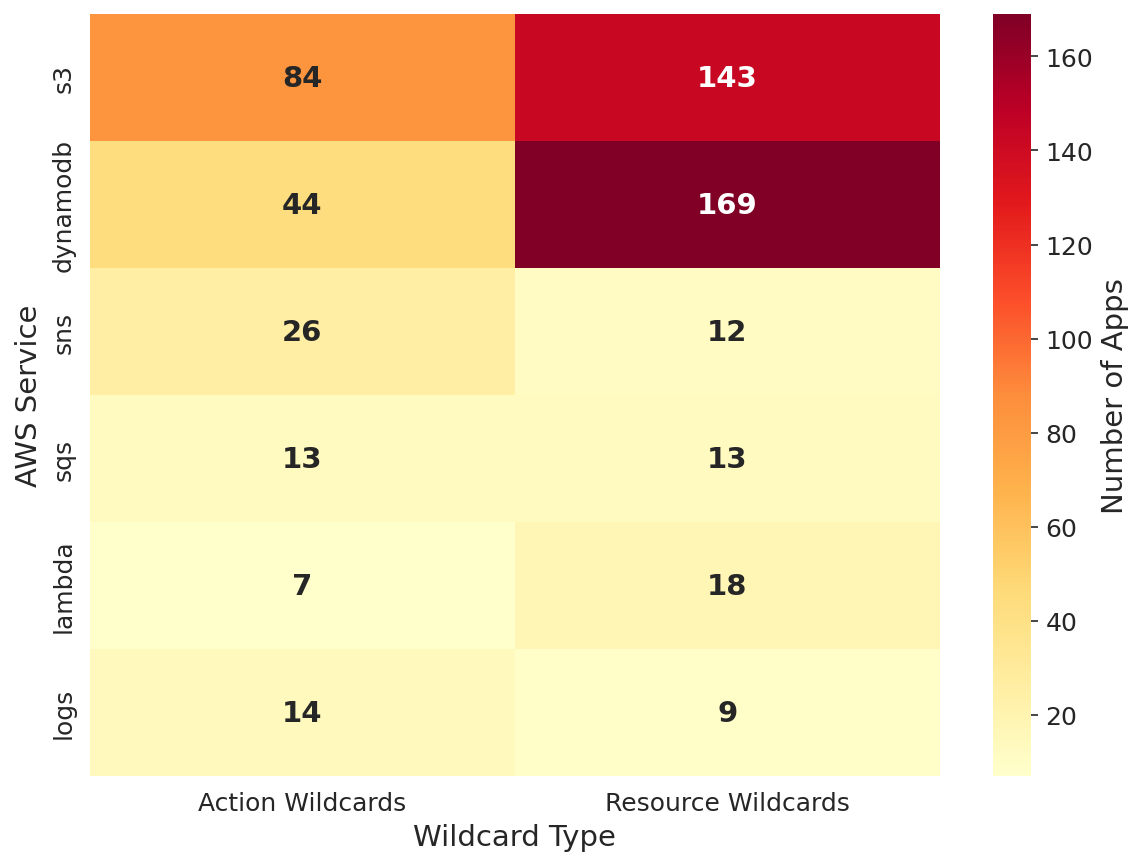}
    \caption{Top 5 AWS service wildcard usage. S3 and DynamoDB services have high wildcard usage patterns across permissions and resources.}
    \label{fig:aws-service-wildcards}
\end{figure}

\shortsectionBf{Wildcard Co-occurrence.} Over-privilege reaches its peak when permission and resource wildcards combine (compounded over-privilege), amplifying the attack surface multiplicatively. In our dataset, 21\% (166 apps) exhibited this dangerous combination. Most alarmingly, 7 applications configured the worst possible scenario: full permission wildcards (\texttt{*}) paired with full resource wildcards (\texttt{Resource: "*"}), granting functions unrestricted access to the entire AWS account. Against such configurations, even trivial vulnerabilities such as an SQL injection flaw, an SSRF bug, or a compromised dependency, become account-takeover vectors. An attacker could read all data across all services, create privileged IAM credentials, delete production databases, or pivot to other AWS accounts via cross-account roles with no technical barriers.

Beyond these extreme cases, partial wildcard combinations remain highly dangerous. A policy that combines service-level action wildcards (e.g., \texttt{s3:*}) with full resource wildcards (\texttt{Resource: "*"}) grants all S3 operations across all buckets, enabling massive data exfiltration. Similarly, prefix action wildcards (e.g., \texttt{dynamodb:Batch*}) paired with full resource wildcards permit all batch operations on every DynamoDB table. Even seemingly "reasonable" combinations of service-scoped resources with broad action wildcards remove the containment boundaries that should limit the blast radius of a compromised function. Table~\ref{tab:wildcard-permutations} enumerates the distribution of these dangerous co-occurrence patterns across our dataset.

\begin{table}[h]
\centering
\caption{Wildcard Permutation Analysis (Action and Resource Wildcards)}
\label{tab:wildcard-permutations}
\setlength{\tabcolsep}{6pt}
\renewcommand{\arraystretch}{1.2}
\begin{tabular}{lcc}
\hline
\textbf{Wildcard Combination} & \textbf{Count(\%)} \\
\hline
Full Action + Full Resource         & 7 (4\%)  \\
Full Action + Service Resource      & 1 (0.6\%)  \\
Full Action + Prefix Resource       & 2 (1.2\%)  \\
Service Action + Full Resource      & 114 (68.7\%) \\
Service Action + Service Resource   & 25 (15.1\%) \\
Service Action + Prefix Resource    & 57 (34.3\%) \\
Prefix Action + Full Resource       & 16 (9.6\%) \\
Prefix Action + Service Resource    & 2 (1.2\%)  \\
Prefix Action + Prefix Resource     & 8 (4.8\%)  \\
\hline
\end{tabular}
\end{table}

\shortsectionBf{Wildcard usage across Services.} We find that wildcard usage concentrates most heavily on services that store sensitive data, amplifying data breach risks. As shown in Figure~\ref{fig:aws-service-wildcards}, S3 and DynamoDB, the primary data storage services in AWS serverless architectures, exhibit the highest wildcard usage with 312 applications defining resource wildcards and 128 applications using permission wildcards across these services.

\shortsectionBf{Attack Surface Approximation.}
We characterize the attack surface and potential blast radius of a compromised function by 
mapping permissions to a subset of the MITRE ATT\&CK framework~\cite{mitre}. Using keyword-based heuristics (Appendix~\ref{app:permission-classification}), we classify each permission into one of eight categories based on its potential role in an attack:
\texttt{Reconnaissance}, 
\texttt{Data Exfiltration}, 
\texttt{Credential Access}, 
\texttt{Privilege Escalation}, 
\texttt{Data Tampering}, 
\texttt{Data Destruction}, 
\texttt{Denial of Service}, 
\texttt{Resource Hijacking}, 
and \texttt{Defense Evasion}. 

We observe that many applications have permissions that can enable full attack chains. Data Tampering permissions, which allow attackers to change records, add harmful content, or alter how the app works, are found in 76.6\% of applications. Data Exfiltration permissions are present in 59.9\% of applications. 41.3\% of applications have Data Destruction permissions that could be used for ransomware attacks or sabotage, and 21.4\% have Privilege Escalation permissions that allow attackers to assume higher-privileged roles. Credential Access permissions, which allow attackers to create persistent backdoors using new API keys or session tokens, are found in 9.0\% of applications. This spread of permissions means that if a function is compromised, attackers often do not need to use multiple exploits as they already have what they need to carry out multi-stage attacks. These can include mapping resources, stealing data, maintaining access, and hiding their actions. The high number of applications with destructive permissions (41.3\% can delete data) is especially worrying, as it means nearly half of serverless applications could be used to destroy data or mount denial-of-service attacks if any function is compromised.

\begin{figure}[t]
    \centering
    \includegraphics[width=\linewidth]{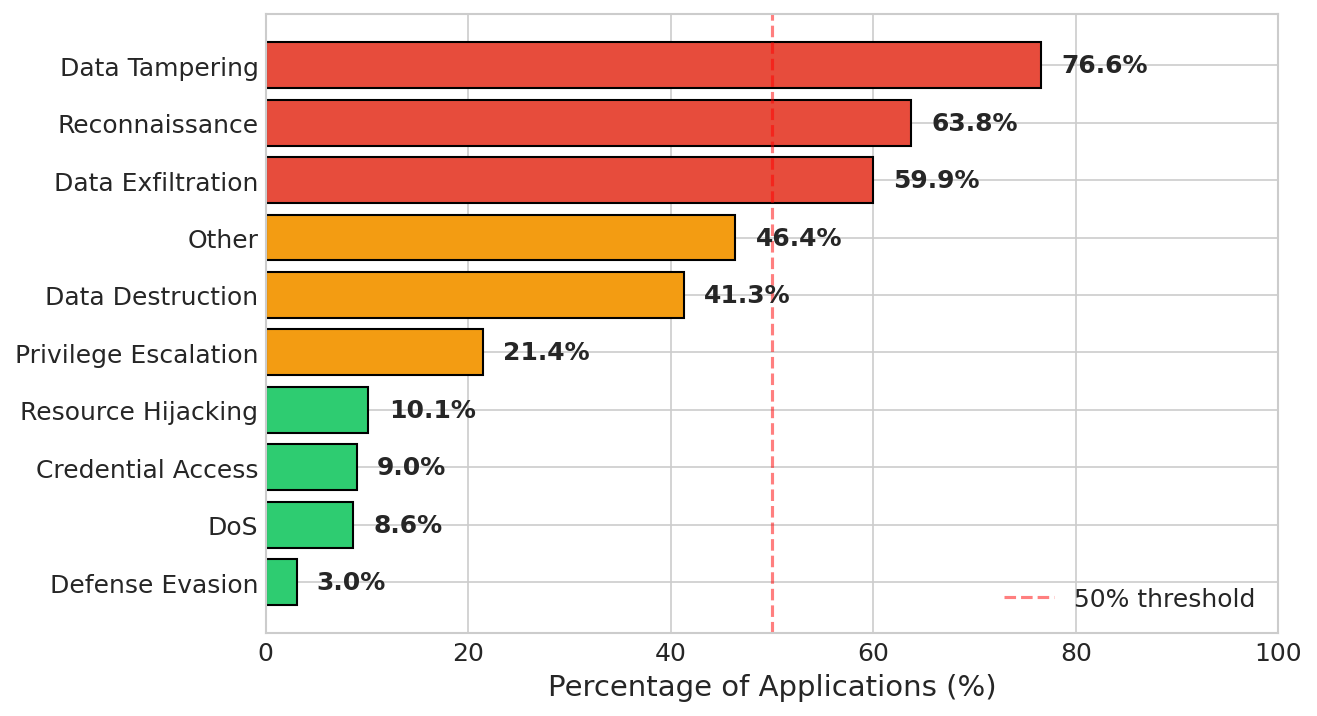}
    \caption{Distribution of Permission Type. Most assigned permissions (76.6\%)  allow functions to manipulate data. 21.4\% allow privilege escalation, and 41.3\% allow the destruction of data or storage bases.}
    \label{fig:perm_classification_app_prevalence}
\end{figure}

\begin{figure*}[t]
    \centering
    \begin{subfigure}[b]{0.48\linewidth}
        \centering
        \includegraphics[width=0.8\linewidth]{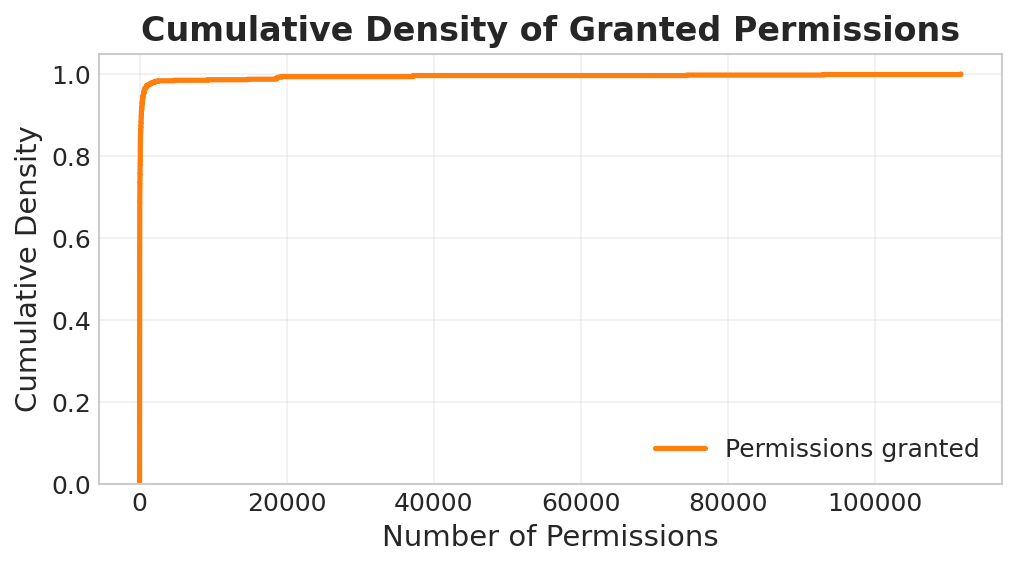}
        \caption{Granted Permissions Density}
        \label{fig:app_dens_a}
    \end{subfigure}
    \hfill
    \begin{subfigure}[b]{0.48\linewidth}
        \centering
        \includegraphics[width=0.8\linewidth]{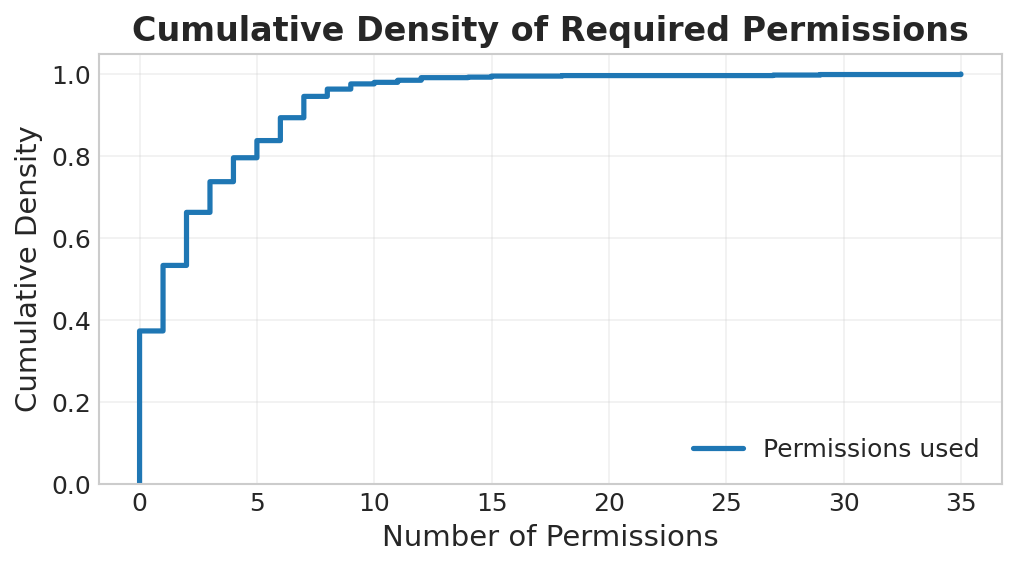}
        \caption{Required Permissions Density}
        \label{fig:app_dens_b}
    \end{subfigure}
    \caption{Cumulative density of permissions required against density of permissions granted per application. 76.7\% of applications require four or fewer permissions to function. However, 65.3\% are granted five or more}
    \label{fig:app_dens}
\end{figure*}

\begin{figure*}[t]
    \centering
    \begin{subfigure}[b]{0.48\linewidth}
        \centering
        \includegraphics[width=0.8\linewidth]{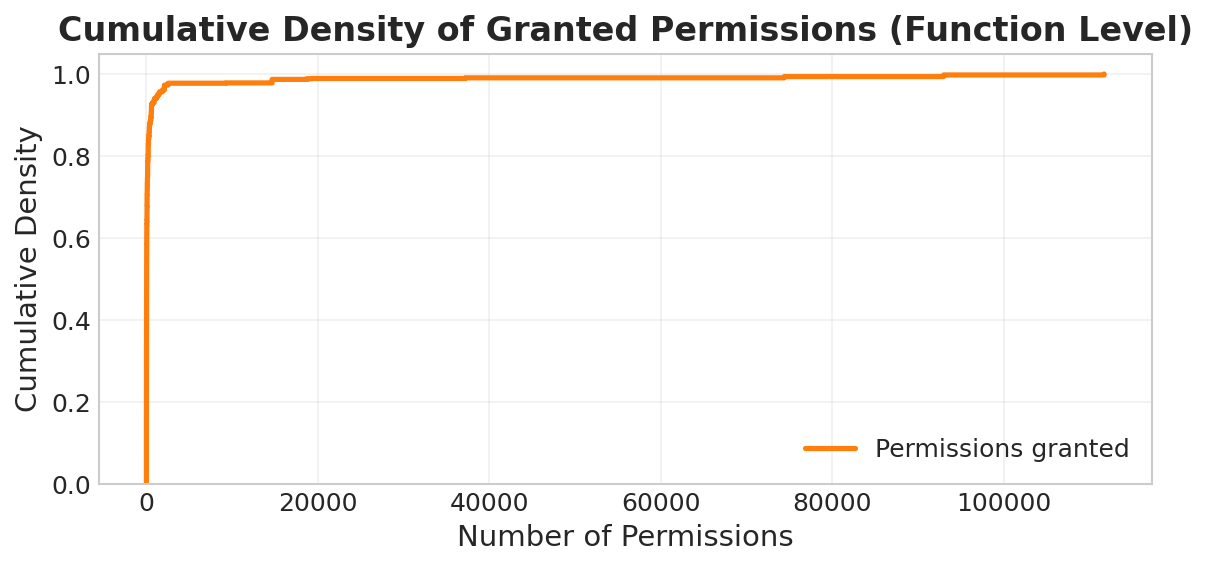}
        \caption{Granted Permissions Density}
        \label{fig:func_dens_a}
    \end{subfigure}
    \hfill
    \begin{subfigure}[b]{0.48\linewidth}
        \centering
        \includegraphics[width=0.8\linewidth]{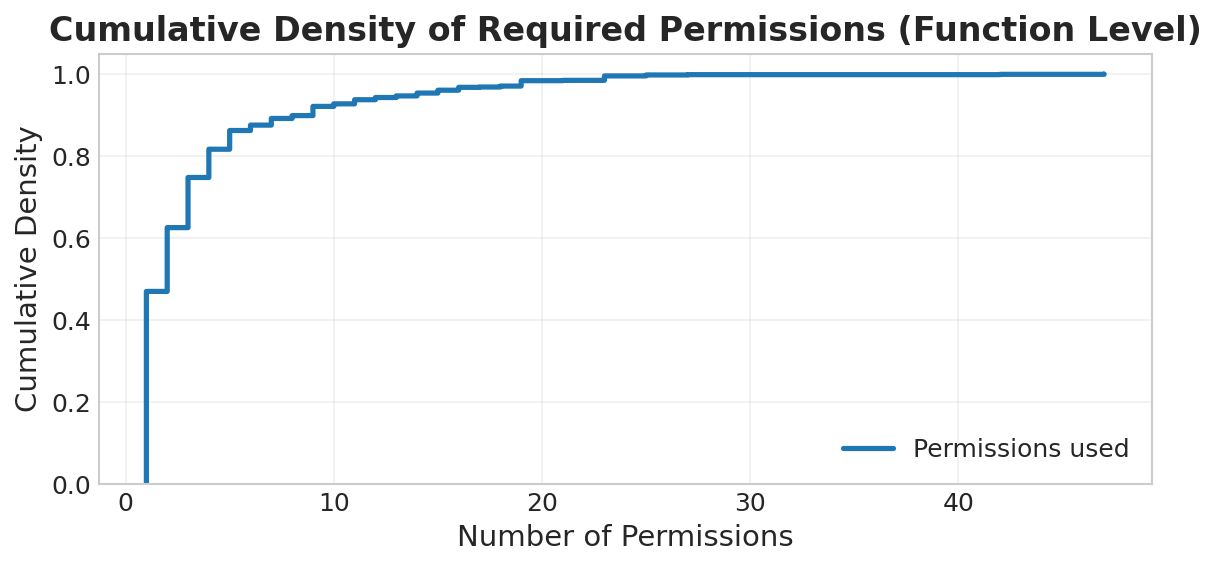}
        \caption{Required Permissions Density}
        \label{fig:func_dens_b}
    \end{subfigure}
    \caption{Cumulative density of permissions required against cumulative density of permissions granted per function. 46.7\% of the applications require two or fewer permissions. However, 87.7\% were granted three or more permissions}
    \label{fig:func_dens}
\end{figure*}

\subsection{PrivLess: Over-privilege Analysis}
\shortsectionBf{Permission Density and Privilege Reduction.}
We measured the gap between permissions granted in developer-defined IAM policies and the minimal set \pName infers from function interactions. Across the 789 applications, we observe a striking disparity between granted and required permissions. Developers collectively granted 538{,}390 effective permissions (after wildcard expansion) against 1{,}896 required permissions determined by \pName. This represents a staggering 99.65\% privilege reduction potential. The mean granted was 683.37 versus 2.40 required (285$\times$), with the disparity holding at the median where applications were granted 9 permissions while requiring only 1. This shows that over-privilege is not limited to outliers but is a systemic issue. In total, 47.7\% of applications (376) carry more permissions than \pName determines are necessary. The CDF in Figure~\ref{fig:app_dens} illustrates the systematic mismatch: 76.7\% of applications require four or fewer permissions to function, yet 65.3\% are granted five or more. Only 26.1\% of applications receive permissions closely matching their actual needs. Moreover, at the function level (Figure~\ref{fig:func_dens}), 46.7\% of functions required two or fewer permissions to function but 87.7\% were granted more, demonstrating that permission grants are decoupled from code requirements.

\shortsectionBf{Wildcard Influence on Over-privilege.}
Wildcards significantly amplify over-privilege. As shown in Table~\ref{tab:permission-analysis}, applications with wildcards exhibit a 98.69\% privilege reduction potential compared to 61.78\% for those without, and are over-privileged at higher rates (59.2\% vs. 44.1\%). This reflects how unrestricted action patterns compound permission sprawl.

\shortsectionBf{Policy Type Influence on Overprivilege.}
Similarly, policy scope choice directly influences overprivilege severity(Table~\ref{tab:permission-analysis}). Global-only policies, used in 699 applications (88.6\%), had a 73.97\% privilege reduction potential. Function-level policies, though used by only 52 applications, demonstrate substantially lower over-privilege with a 41.35\% reduction potential. This affirms that function-based policies minimize the blast radius, but do not automatically eliminate overprivilege and require careful privilege assignment. Hybrid policies (both global and function-level) split the difference at 88.20\% reduction, suggesting that per-function refinement, even when paired with a global baseline, meaningfully constrains excess permissions. The dominance of global-only policies thus directly drives the systemic over-privilege observed across the dataset.

\begin{table}[h]
\centering
\caption{Permission Usage by Wildcard Presence and Policy Scope}
\label{tab:permission-analysis}
\setlength{\tabcolsep}{6pt}
\renewcommand{\arraystretch}{1.2}
\begin{tabular}{lccc}
\hline
\textbf{Category} & \textbf{Apps} & \textbf{Avg Granted} & \textbf{Reduction} \\
\hline
No Wildcards      & 605 & 14.72    & 61.78\% \\
Has Wildcards     & 184 & 2{,}877.62 & 98.69\% \\
\hline
Global Only       & 699 & 738.89   & 73.97\% \\
Function Only     & 52  & 23.23    & 41.35\% \\
Both              & 19  & 1{,}089.47 & 88.20\% \\
\hline
\end{tabular}
\end{table}

\shortsectionBf{Security Impact of Over-Privilege by Threat Category.}
We classified all granted and required IAM permissions into nine threat categories (Reconnaissance, Data Exfiltration, Credential Access, Privilege Escalation, Data Tampering, Data Destruction, DoS, Resource Hijacking, Defense Evasion) to assess real-world risk. The granted vs. required gap is stark: Reconnaissance permissions appear in 48.8\% of applications despite rarely being functionally necessary; Privilege Escalation permissions are granted in 93 applications (18.8\%) but required in only 19; most critically, Defense Evasion permissions (e.g., disabling CloudTrail, deleting CloudWatch alarms) are granted in 12 applications yet required by zero, creating silent persistence paths for attackers. Only Credential Access shows restrained over-granting (77.0\% reduction), suggesting developers exercise discipline when managing secrets but not across other threat categories.

\subsection{Performance Evaluation}
For the end-to-end runtime of our implementation of \pName, we observed a mean runtime of 193.6 seconds across all 789 applications. CodeQL operations dominate with 99.9\% of total execution time consumed by database creation ($\sim24\%$) and query execution ($\sim75\%$). The core Python-side analysis (resource resolution, call-graph scoping, and permission mapping) completes in under 100 ms on average and is never a bottleneck. We noticed a high variance in runtime (p95 $\approx$ 8 min) which reflects differences in codebase size, dependency depth, and language complexity (e.g., JS/TS transpilation), not \pName's algorithm. We argue that \pName's analysis is not in the critical path and operates offline; hence, the seemingly high end-to-end time cost is acceptable. Moreover, alternative static analysis tools (such as pure AST-based analyzers) could eliminate database creation overhead and reduce runtime. We chose CodeQL for its rich query language and relational database semantics, a performance tradeoff justified by enhanced analysis capability and precision.

\section{Case Studies}
\label{sec:case-studies}

To ground the aggregate findings, we examine five applications that collectively span the dominant patterns of over-privilege observed in our dataset (Table~\ref{tab:casestudy-summary}; detailed per-function breakdown in Appendix~\ref{app:case-studies}). For ethical reasons, we anonymized the case-study apps.

The five applications exhibit four patterns: \emph{over-scoped service wildcards}, where \texttt{s3:*} (229 actions) and \texttt{rekognition:*} (76 actions) are applied globally in \texttt{Anonymized-1} (3{,}110 effective permissions) when one specific action per service suffices; \emph{speculative service inclusion} (service suites declared for integrations never implemented or removed, as in \texttt{Anonymized-2}'s unused Rekognition suite — 126 grants, zero calls); \emph{deployment tooling in runtime roles} (\texttt{codedeploy:*} alone accounts for 517 of 570 grants in \texttt{Anonymized-4}); and \emph{over-declared operations} (a global policy in the \texttt{Anonymized-3\&5} app grants all anticipated actions uniformly to every function regardless of individual needs). Reductions range from 83\% to 99.8\%, showing that over-privilege scales with policy scope rather than application complexity.

\begin{table}[h]
\centering
\caption{Case Study: Privilege Reduction}
\label{tab:casestudy-summary}
\setlength{\tabcolsep}{5pt}
\renewcommand{\arraystretch}{1.2}
\begin{tabular}{lcrrrr}
\hline
\textbf{App} & \textbf{Funcs} & \textbf{Grant} & \textbf{Req} & \textbf{Reduction} \\
\hline
Anonymized-1  & 10 & 3{,}110 &  7 & 99.8\%  \\
Anonymized-2            & 18 &    522 & 22 & 95.8\% \\
Anonymized-3    & 20 &     82 & 11 & 86.6\%  \\
Anonymized-4  & 11 &    570 &  5 & 99.1\% \\
Anonymized-5   &  5 &     30 &  5 & 83.3\% \\
\hline
\end{tabular}
\end{table}

\shortsectionBf{Functionality Validation.}
Reduced permissions are only useful if the application still works as intended: if PrivLess breaks a function, that regression would undermine both PrivLess and our measurements.

For each case-study application, we deployed the baseline-policy version and the PrivLess-reduced-policy version in LocalStack, a local AWS emulator that we configure to enforce IAM permissions~\cite{localstack}, and on real AWS. We invoked each function under identical inputs in both versions and compared results (Table~\ref{tab:func-detail}). We also ran \emph{negative controls}: calls to API actions PrivLess removed from the policy, which should fail if the reduction was correct. Since each function implements one well-defined task, invoking its handler gives full code coverage.

Across the five applications, PrivLess preserved baseline functionality in 29/31 (94\%) functions on LocalStack and 21/31 (68\%) functions on AWS. Functionality is ``preserved'' if PrivLess's version passes when the baseline passes, or fails the same way when the baseline fails. Only one case is a genuine regression, caused by a static-analysis blind spot in a third-party middleware call; a second, superficially similar failure is an unrelated flake from randomized application logic, not caused by the reduced policy. Every negative control we ran was correctly denied. Cases excluded from the totals above were blocked by environment or deployment issues unrelated to PrivLess's .

\begin{table}[h!]
\centering
\caption{Functionality validation detail, by application, function, and environment.}
\label{tab:func-detail}
\setlength{\tabcolsep}{4pt}
\begin{tabular}{l cc cc}
\toprule
& \multicolumn{2}{c}{\textbf{LocalStack}} & \multicolumn{2}{c}{\textbf{AWS}} \\
\cmidrule(lr){2-3} \cmidrule(lr){4-5}
\textbf{App/Function} & \textbf{Base.} & \textbf{\pName} & \textbf{Base.} & \textbf{\pName} \\
\midrule

\multicolumn{5}{l}{\textbf{Anonymized-1}} \\
homepage                                            & \Pass & \Pass & \Pass & \Pass \\
features-list                                       & \Pass & \Pass & \Pass & \Pass \\
create-snippet                                      & \Pass & \Pass & \Pass & \Pass \\
list-user-snippets                                  & \Pass & \Pass & \Pass & \Pass \\
get-snippet                                         & \Pass & \Pass & \Pass & \Pass \\
update-snippet                                      & \Pass & \Pass & \Pass & \Pass \\
delete-snippet                                      & \Pass & \Pass & \Fail & \Fail \\
upload                          & \Blocked & \Blocked & \Fail & \Fail \\
negative controls & \NA & \Pass & \NA & \Pass \\
\midrule

\multicolumn{5}{l}{\textbf{Anonymized-2}} \\
Failed to Deploy    & \Blocked & \Blocked & \Blocked & \Blocked \\
\midrule

\multicolumn{5}{l}{\textbf{Anonymized-3}} \\
confirm-user-signup                                 & \Pass & \Pass & \Blocked & \Blocked \\
get-image-upload-url                                & \Pass & \Pass & \Blocked & \Blocked \\
tweet                                               & \Pass & \Pass & \Blocked & \Blocked \\
retweet                                             & \Pass & \Pass & \Blocked & \Blocked \\
reply                                               & \Pass & \Pass & \Blocked & \Blocked \\
get-tweet-creator                                   & \Pass & \Pass & \Blocked & \Blocked \\
send-direct-message                                 & \Pass & \Pass & \Blocked & \Blocked \\
search                                & \Fail & \Fail & \Blocked & \Blocked \\
negative controls                               & \Pass & \Pass & \Blocked & \Blocked \\
\midrule

\multicolumn{5}{l}{\textbf{Anonymized-4}} \\
get-index                                           & \Pass & \Fail & \Blocked & \Blocked \\
get-restaurants                       & \Fail & \Fail & \Fail & \Fail \\
search-restaurants                                  & \Pass & \Pass & \Pass & \Pass \\
place-order                                         & \Pass & \Pass & \Pass & \Pass \\
place-order-triggers             & \Pass & \Pass & \Pass & \Pass \\
accept-order                                        & \Fail & \Fail & \Fail & \Fail \\
fulfill-order                                       & \Fail & \Fail & \Fail & \Fail \\
retry-notify-restaurant                & \Pass & \Pass & \Pass & \Fail \\
retry-notify-user                      & \Fail & \Fail & \Pass & \Pass \\
auto-create-api-alarms                              & \Fail & \Fail & \Fail & \Fail \\
negative controls                                   & \NA & \Pass & \NA & \Pass \\
\midrule

\multicolumn{5}{l}{\textbf{Anonymized-5}} \\
create                                             & \Pass & \Pass & \Pass & \Pass \\
list                                               & \Pass & \Pass & \Pass & \Pass \\
get                                                 & \Pass & \Pass & \Pass & \Pass \\
update                                              & \Pass & \Pass & \Pass & \Pass \\
delete                                              & \Pass & \Pass & \Pass & \Pass \\
negative controls & \NA & \Pass & \NA & \Pass \\
\bottomrule
\end{tabular}

\vspace{2pt}

\parbox{\columnwidth}{%
\scriptsize
\textit{Legend:} \Pass\ Pass; \Fail\ Fail; 
\Blocked\ deployment blocker; \NA\ not applicable.
}
\end{table}

\section{Limitations}



\shortsectionBf{SDK-based tracing.} Our tracing has two blind spots: (i) it misses calls routed through third-party SDK wrappers (e.g., \texttt{dynamoose}) that do not go through the instrumented client, and (ii) it cannot observe direct HTTP calls or raw API invocations that bypass the SDK entirely. In both cases, PrivLess may miss the corresponding permissions, resulting in an under-provisioned policy.

\shortsectionBf{Knowledge base fidelity.} No cloud provider publishes a comprehensive authoritative, single-source mapping from API actions to required permissions, so our knowledge base is necessarily best-effort. We mitigate this by favoring conservative permission mappings and by drawing exclusively from official provider documentation and resources. We encourage providers, who have direct oversight of their services, to publish and maintain such a mapping as a canonical source of truth.

\section{Related Works}
\textit{Cloud IAM Auditing Tools:}
AWS Zelkova\cite{zelkova-blog, bbb2020} uses SMT solving to evaluate policy conditions, while AWS IAM Access Analyzer\cite{iam-access-analyzer}, Azure Monitor\cite{AzureMonitor}, and Google Security Command Center\cite{googleSecurityCommand} provide post-deployment auditing and visibility. These tools are \textit{reactive}—reporting overprivilege after deployment. PrivLess enables \textit{pre-deployment} measurement of the permission gap at scale by analyzing source code directly.

\textit{Information Flow Control:}
Systems like Valve~\cite{Valve}, Trapeze~\cite{Trapeze}, and SecLambda~\cite{SecLambda} enforce data confidentiality in serverless applications but assume IAM policies are already correctly scoped. PrivLess addresses the upstream problem: quantifying the extent to which real-world policies exceed what functions actually require.

\textit{Static Policy Analysis:}
GRASP\cite{GRASP} constructs reachability graphs from configurations, AutoArmor\cite{autoarmor} analyzes microservice patterns, and Will.IAM\cite{willIAm} enforces access control dynamically. These rely on configuration correctness or runtime instrumentation. ALPS\cite{shin2026alps}, the closest related work, applies static analysis to infer minimum permissions and enforces them via runtime hooks across multiple cloud providers. Unlike ALPS, which targets \textit{enforcement}, PrivLess targets \textit{measurement}—quantifying overprivilege in large real-world corpora without code modification or runtime instrumentation. Growlithe\cite{gupta2024growlithe} combines static and runtime approaches but requires policy annotations. D'Antoni et al.\cite{DAntoni2024} infer policies from request logs, but log-based approaches miss untriggered code paths.

\section{Conclusion}

We present a systematic overprivilege analysis framework, \pName, that quantifies the gap(overprivilege) between developer-defined IAM policies and actual applications' IAM needs. Applied to 789 real-world serverless applications, our study establishes that overprivilege is endemic: 47.7\% of applications carry excess permissions with an aggregate privilege reduction potential of upto 99.65\%. Wildcard policies are the primary driver, inflating overprivilege to 799$\times$ for affected applications (274$\times$ higher than those without wildcards). These excess permissions pose concrete risks, as 18.8\% of applications hold unnecessary privilege escalation capabilities, and 12 applications grant defense evasion permissions they never use. These findings demonstrate that manual IAM management has systematically failed in serverless environments, making automated analysis not merely beneficial but essential infrastructure for understanding and remediating overprivilege at scale.







\bibliographystyle{plainurl}
\bibliography{bibliography}

\appendix
\section{Appendix}



\section*{Ethical Considerations}
Our dataset consists of publicly available open-source repositories. While this data is publicly accessible, our analysis reveals security vulnerabilities (overprivileged policies) that developers may not have been aware of. To respect developer privacy and intent, we anonymize applications in aggregate findings and encourage repository maintainers to use PrivLess to audit their own applications.

\shortsectionBf{Responsible Disclosure of Case Studies:}
The five applications selected for detailed case studies are currently anonymized in this manuscript. We conducted responsible disclosure by directly emailing repository owners with detailed findings, security implications, and remediation recommendations on June 2, 2026. As at now, we have received no response back from the repository owners. We deliberately refrained from using public channels (pull requests or GitHub issues) for disclosure, as publicly exposing overprivilege vulnerabilities could enable attackers to exploit these weaknesses before fixes are deployed. This private disclosure approach allows developers to address issues on their own timeline without creating attack opportunities. If repository owners provide feedback, explicit consent, and evidence of fixes before publication, we commit to de-anonymizing those applications and crediting the maintainers for their proactive response. This practice balances transparency, security, and developer agency.

\section*{Open Science}
To support reproducibility and encourage further research, we make our artifacts publicly available. Our submission includes:

\textit{Source Code and Tools:} The complete implementation of our system, \pName is available at https://anonymous.4open.science/r/privLess-framework-21F4/.

\textit{Case Study Applications:} A curated dataset of our serverless applications with security policies collected from public GitHub repositories, including extracted IAM policies for the case study application, required permissions, and computed metrics. Due to responsible disclosure commitments, the dataset is initially released in anonymized form. Following notification of affected repository owners and appropriate remediation periods, we will release the deanonymized dataset to enable full reproducibility.

\textit{Experimental Artifacts:} All scripts used for data collection, analysis, and visualization of results presented in this paper, along with configuration files for replicating our experimental setup.

\textit{Documentation:} Each artifact includes comprehensive documentation detailing system requirements, installation procedures, and usage instructions. Researchers can reproduce our complete analysis pipeline or apply our tools to new serverless applications. Detailed instructions are provided in the README files accompanying each component.

\subsection{Case Studies - Detailed Analysis}
\label{app:case-studies}

\shortsectionBf{Anonymized-1 App:} is a code-sharing platform that allows users to create, view, update, and delete code snippets, with image upload support. Its 10 Lambda functions are: \texttt{verify} (domain ownership verification), \texttt{main} (homepage), \texttt{features} (feature listing), \texttt{create} (snippet creation), \texttt{login} (WeChat OAuth), \texttt{userCode} (list snippets by user), \texttt{code} (get snippet detail), \texttt{updateCode} (edit snippet), \texttt{deleteCode} (remove snippet), and \texttt{upload} (image upload with text recognition).

The global IAM policy carries three statements applied uniformly to all 10 functions: six specific DynamoDB actions (\texttt{GetItem}, \texttt{PutItem}, \texttt{UpdateItem}, \texttt{Query}, \texttt{Scan}, \texttt{DeleteItem}) on the application table; \texttt{s3:*} on the upload bucket; and \texttt{rekognition:*} on all resources. After wildcard expansion using the permission model (229 S3 actions, 76 Rekognition actions), the effective total is $3{,}110$ permissions across the 10 functions.

\pName detects service calls in 6 of the 10 functions. The function handlers (\texttt{code}, \texttt{userCode}, \texttt{updateCode}, \texttt{deleteCode}, \texttt{create}) use four DynamoDB operations (\texttt{Scan}, \texttt{PutItem}, \texttt{UpdateItem}, \texttt{DeleteItem}); the relaxer augments \texttt{GetItem} and \texttt{Query} back since both appear in the developer policy and DynamoDB is a directly-detected service. The \texttt{upload} handler uses two services: \texttt{s3:PutObject} (direct call) and \texttt{rekognition:DetectText}, which is called through the local utility module \texttt{imageAnalyser.js} using Rekognition's nested image schema (\texttt{\{\ Image:\ \{\ S3Object:\ \{\ Bucket,\ Name\ \}\ \}\ \}}). The extractor recognises \texttt{Image} as a resource parameter and correctly attributes the call to \texttt{upload}. The inferred minimum policy requires 8 unique permissions — six DynamoDB actions, \texttt{s3:PutObject}, and \texttt{rekognition:DetectText} — yielding a 99.7\% reduction ($3{,}110$ to 8).

The case illustrates how service-level wildcards compound over-privilege across two dimensions. The \texttt{s3:*} wildcard grants all 229 S3 actions to every function when only \texttt{upload} needs \texttt{PutObject}; \texttt{rekognition:*} grants all 76 Rekognition actions globally for what is ultimately a single \texttt{DetectText} call in one function. In both cases \pName identifies the precise minimal action and produces per-function policies that eliminate the wildcard entirely.

\shortsectionBf{Anonymized-2 App:} is a photo-management platform that stores images in S3, manages user metadata in DynamoDB, and identifies faces using Amazon Rekognition. It has 18 Lambda functions that handle the full lifecycle: \texttt{create}, \texttt{get}, \texttt{query}, \texttt{update}, \texttt{delete}, \texttt{thumbs}, \texttt{personThumb}, \texttt{fotoEvents}, \texttt{indexes}, \texttt{indexesUpdate}, \texttt{faces}, \texttt{people}, \texttt{peopleUpdate}, \texttt{peopleMerge}, \texttt{person}, \texttt{stream}, \texttt{collectionCreate}, and \texttt{collectionDelete}.

The global IAM policy lists 29 permission entries applied uniformly to all 18 functions — 12 DynamoDB operations, four S3 actions across two buckets, two Lambda actions, four CloudWatch Logs actions, and a seven-action Rekognition suite (\texttt{IndexFaces}, \texttt{CreateCollection}, \texttt{DeleteCollection}, \texttt{DeleteFaces}, \texttt{DetectLabels}, \texttt{ListFaces}, \texttt{SearchFaces}) — totalling 522 declared permissions across all functions. No wildcards appear, so effective equals declared.

\pName traces SDK calls across all 18 functions and detects no code path that invokes any Rekognition action: the seven-action suite (126 grants across 18 functions) exists entirely as policy bloat. The application's authentication module uses the Laconia dependency-injection framework, which passes AWS SDK client objects as callback parameters; this pattern is opaque to static analysis, so Cognito calls in the auth layer are not attributed to Lambda handlers. Since the developer's IAM policy likewise grants no Cognito permissions to the Lambda execution role, this analysis gap does not affect the output. And because the app uses a single global IAM policy, the relaxer assigns the same 22-permission set (the developer's 29 actions minus the seven Rekognition actions) to every function, yielding a $24\times$ ratio and 95.8\% reduction potential (522 to 22 unique permissions).

\shortsectionBf{Anonymized-3 App:} is a Twitter-clone social backend built on AWS AppSync and DynamoDB. Unlike the previous applications, it assigns function-level IAM policies to each of its 20 Lambda functions, which implement user management (\texttt{confirmUserSignup}, \texttt{getImageUploadUrl}), social graph operations (\texttt{tweet}, \texttt{retweet}, \texttt{unretweet}, \texttt{reply}), follower distribution (\texttt{distributeTweets}, \texttt{distributeTweetsToFollower}), search and discovery (\texttt{search}, \texttt{getHashTag}, \texttt{syncUsersToAlgolia}, \texttt{syncTweetsToAlgolia}), notifications (\texttt{notify}, \texttt{notifyLiked}, \texttt{notifyDmed}), messaging (\texttt{sendDirectMessage}, \texttt{getTweetCreator}), and infrastructure management (\texttt{firehoseTransformer}, \texttt{setResolverLogLevelToAll}, \texttt{setResolverLogLevelToError}). Most functions are granted individual DynamoDB operations against named table ARNs; search-related functions additionally receive \texttt{ssm:GetParameters} for Algolia credentials; and a global policy applies \texttt{xray:PutTraceSegments} and \texttt{xray:PutTelemetryRecords} to all 20 functions. After summing across all functions, the developer declares 82 total effective permissions.

\pName identifies 11 unique actions in actual use: six DynamoDB operations (\texttt{GetItem}, \texttt{PutItem}, \texttt{UpdateItem}, \texttt{DeleteItem}, \texttt{Query}, \texttt{BatchWriteItem}), two S3 actions (\texttt{PutObject}, \texttt{PutObjectAcl}), \texttt{ssm:GetParameters}, \texttt{iam:PassRole}, and \texttt{dynamodb:BatchGetItem}. The two globally declared X-Ray actions do not appear in any application code path; the relaxer evicts them from the generated policy. The result is a $7.5\times$ ratio and 86.6\% reduction (82 to 11 unique permissions). This case shows that function-level policies substantially narrow the blast radius compared to global grants, but excess still accumulates through operational tooling (X-Ray) declared speculatively across all functions.

\shortsectionBf{Anonymized-4 App:} is a restaurant ordering application comprising 11 functions: \texttt{get-index}, \texttt{get-restaurants}, \texttt{search-restaurants}, \texttt{place-order}, \texttt{notify-restaurant}, \texttt{retry-notify-restaurant}, \texttt{accept-order}, \texttt{notify-user}, \texttt{retry-notify-user}, \texttt{auto-create-api-alarms}, and \texttt{fulfill-order}.

Its global IAM role carries three entries: \texttt{cloudwatch:PutMetricData}, the X-Ray pair, and \texttt{codedeploy:*}. The last is the primary driver of over-privilege: the CodeDeploy service-level wildcard expands to 47 runtime permissions and is applied to all 11 functions, contributing 517 of the 570 total effective permissions. The intent is to support blue/green deployment automation, but CodeDeploy actions are invoked by deployment pipelines, not by Lambda function code at runtime — no function ever calls a CodeDeploy API. Function-level overrides add targeted permissions: \texttt{dynamodb:scan} (restaurants table), \texttt{kinesis:PutRecord} (order events), \texttt{sns:Publish} (notification topics), \texttt{execute-api:Invoke}, \texttt{ssm:GetParameters*} (prefix wildcard expanding to 2 SSM actions), and \texttt{secretsmanager:GetSecretValue}.

\pName traces SDK calls and identifies five unique actions across the 11 functions: \texttt{cloudwatch:PutMetricData} (all functions), \texttt{cloudwatch:PutMetricAlarm} (\texttt{auto-create-api-alarms} only), \texttt{kinesis:PutRecord} (order-processing functions), \texttt{sns:Publish} (notification functions), and \texttt{dynamodb:Scan} (restaurant-query functions). Although Cognito calls appear in test-helper files, the developer's runtime IAM policy does not include Cognito permissions for Lambda; the relaxer correctly evicts those calls, excluding test infrastructure from the generated function policies. No code path invokes CodeDeploy at runtime, confirming that the wildcard's 47-action expansion (517 of the 570 total grants) is pure overhead. The inferred minimum policy requires five permissions — none from CodeDeploy or X-Ray — yielding a $114\times$ ratio and 99.1\% reduction (570 to 5 unique permissions). This case illustrates how deployment tooling bundled into runtime IAM roles produces systematic over-privilege across every function in the service.

\shortsectionBf{Anonymized-5 App:} is a canonical Serverless Framework starter for a CRUD REST API backed by DynamoDB, comprising five Lambda functions each responsible for a single operation: \texttt{create} (POST), \texttt{list} (GET all), \texttt{get} (GET by id), \texttt{update} (PUT), and \texttt{delete} (DELETE). The developer declares a single global IAM policy granting six DynamoDB actions — \texttt{Query}, \texttt{Scan}, \texttt{GetItem}, \texttt{PutItem}, \texttt{UpdateItem}, and \texttt{DeleteItem} — on the application table, which is pplied uniformly to all five functions. With no wildcards, the effective permissions equal declared permissions: 30 total across the functions.

\pName traces one SDK call per function and identifies five distinct DynamoDB actions in use: \texttt{PutItem} (\texttt{create}), \texttt{Scan} (\texttt{list}), \texttt{GetItem} (\texttt{get}), \texttt{UpdateItem} (\texttt{update}), and \texttt{DeleteItem} (\texttt{delete}). No function calls \texttt{query}, making \texttt{dynamodb:Query} the sole superfluous permission — granted to all five functions but exercised by none. The inferred minimum policy requires five unique permissions, yielding a $6\times$ ratio and 83.3\% reduction (30 to 5 unique permissions). Beyond service-level eviction, per-function scoping reveals a deeper opportunity: each function needs exactly one DynamoDB action, yet the global policy grants six to all. \pName's per-function output assigns each function its single required action, reducing total grants from 30 to 5 and eliminating cross-function privilege (a \texttt{create} function that also holds \texttt{DeleteItem} can silently delete records). This case establishes the baseline pattern — over-declared operations in a global policy — that the following applications exhibit at larger scale.

\subsection{Permission Classification Rules}
\label{app:permission-classification}

We classify AWS IAM permissions according to the attack capability they could enable if the principal granted the permission were compromised. The classification uses keyword-based heuristics with explicit service-specific overrides and a fixed priority order. We use the following categories: \texttt{Reconnaissance (R)}, \texttt{Data Exfiltration (DE)}, \texttt{Data Destruction (DD)}, \texttt{Credential Access (CA)}, \texttt{Privilege Escalation (PE)}, \texttt{Data Tampering (DT)}, \texttt{Denial of Service (DoS)}, \texttt{Resource Hijacking (RH)}, and \texttt{Defense Evasion (DEv)}. Permissions matching none of these rules are classified as \texttt{Other}.

Given a permission's AWS service and action, classification proceeds as follows.

\subsubsection*{Service-Specific Overrides}

We first apply the following service-action overrides, which take precedence over all general rules:

\begin{itemize}
\item IAM \texttt{get}/\texttt{list}, Secrets Manager \texttt{describe}/\texttt{list}, SSM \texttt{describe}/\texttt{list}, and KMS \texttt{get}/\texttt{list} $\rightarrow$ \texttt{Reconnaissance}.
\item STS \texttt{assume}/\texttt{get}, and Lambda \texttt{update}/\texttt{create} $\rightarrow$ \texttt{Privilege Escalation}.
\item Secrets Manager \texttt{get}, SSM \texttt{get}, and KMS \texttt{decrypt} $\rightarrow$ \texttt{Credential Access}.
\item Lambda \texttt{invoke} $\rightarrow$ \texttt{Resource Hijacking}.
\item CloudTrail \texttt{stop}/\texttt{delete}, GuardDuty \texttt{delete}, and CloudWatch Logs \texttt{delete} $\rightarrow$ \texttt{Defense Evasion}.
\end{itemize}

\subsubsection*{Priority-Ordered Rules}

If no override applies, we evaluate the following rules in order and assign the first matching category:

\begin{enumerate}
\item \textbf{Credential Access (CA):} A permission is classified as credential access when it retrieves or decrypts sensitive information from Secrets Manager, SSM, or KMS, or when its action contains both a sensitive-data term (e.g., \texttt{secret}, \texttt{password}, \texttt{key}, \texttt{token}, \texttt{credential}, or \texttt{parameter}) and a retrieval operation such as \texttt{get}, \texttt{decrypt}, or \texttt{retrieve}. Listing or describing resources alone does not qualify.

\item \textbf{Privilege Escalation (PE):} A permission is classified as privilege escalation when it modifies IAM, STS, Lambda, or EC2 privilege-related configuration using actions such as \texttt{attach}, \texttt{put}, \texttt{create}, \texttt{update}, \texttt{modify}, \texttt{add}, or \texttt{associate}, together with terms such as \texttt{policy}, \texttt{role}, \texttt{user}, \texttt{group}, \texttt{permission}, \texttt{assume}, or \texttt{pass}. Permissions containing \texttt{PassRole} or \texttt{AssumeRole}, as well as Lambda permissions involving function code, are also classified as privilege escalation.

\item \textbf{Data Destruction (DD):} A permission is classified as data destruction when a destructive action such as \texttt{delete}, \texttt{remove}, \texttt{destroy}, \texttt{terminate}, or \texttt{drop} targets an object or resource such as an object, record, table, bucket, function, instance, or volume, or operates on a destructive data service such as S3, DynamoDB, RDS, Redshift, ElastiCache, or DocumentDB.

\item \textbf{Denial of Service (DoS):} A permission is classified as denial of service when an action such as \texttt{delete}, \texttt{stop}, \texttt{terminate}, or \texttt{disable} targets a service resource whose disruption can affect availability, including functions, instances, clusters, services, tables, distributions, load balancers, databases, queues, and topics.

\item \textbf{Resource Hijacking (RH):} A permission is classified as resource hijacking when it can execute or launch compute workloads through actions such as \texttt{run}, \texttt{invoke}, \texttt{create}, \texttt{start}, \texttt{execute}, or \texttt{launch} on services including EC2, Lambda, SageMaker, Batch, ECS, or EKS.

\item \textbf{Data Exfiltration (DE):} A permission is classified as data exfiltration when it retrieves, reads, downloads, exports, or fetches data from storage services such as S3, DynamoDB, RDS, Redshift, or ElastiCache, or targets data-bearing resources such as objects, records, files, documents, content, or backups.

\item \textbf{Data Tampering (DT):} A permission is classified as data tampering when it performs a modification operation such as \texttt{put}, \texttt{update}, \texttt{modify}, \texttt{write}, \texttt{edit}, \texttt{change}, \texttt{replace}, \texttt{upload}, or \texttt{insert}, provided that it does not already satisfy the privilege-escalation rule.

\item \textbf{Reconnaissance (R):} A permission is classified as reconnaissance when it performs an informational operation such as \texttt{list}, \texttt{describe}, \texttt{get}, \texttt{scan}, \texttt{query}, or \texttt{search}, provided that it does not target credentials or data-bearing resources as defined above.

\item \textbf{Other (O):} Permissions that match none of the preceding rules are classified as \texttt{Other}.
```

\end{enumerate}

For all rules, keyword matching is based on substring containment within the normalized AWS service and action names (e.g., \texttt{get} matches \texttt{getObject}). The service-specific overrides are evaluated first; otherwise, the first matching rule in the priority order determines the final classification.

\end{document}